\documentclass[aps,11pt,english,tightenlinesletterpaper,superscriptaddress,secnumarabic,nofootinbib]{revtex4}

\pdfoutput=1

\usepackage{multirow}
\usepackage{mypackages}
\usepackage[sort&compress]{natbib}
\usepackage{hyperref} 	
\usepackage[all]{hypcap}  
\hypersetup{
    colorlinks=true,		
    linkcolor=red,		
    citecolor=blue,		
    filecolor=magenta,	
    urlcolor=blue		
}

\usepackage{myterms}
\newcommand{\SM}{\text{\sc sm}}
\newcommand{\HS}{\text{\sc hs}}
\newcommand{\CC}{\text{\sc cc}}
\newcommand{\DE}{\text{\sc de}}
\newcommand{\CW}{\text{\sc cw}}

\begin{document}

\title{Metastability of the False Vacuum in a Higgs-Seesaw Model of Dark Energy}

\author{Lawrence M. Krauss}
\email{krauss@asu.edu}
\affiliation{\small Department of Physics and School of Earth and Space Exploration \\ Arizona State University, Tempe, AZ 85827-1404}
\affiliation{\small Research School of Astronomy and Astrophysics, Mt. Stromlo Observatory, \\ Australian National University, Canberra, Australia 2611}

\author{Andrew J. Long}
\email{andrewjlong@asu.edu}
\affiliation{\small Department of Physics and School of Earth and Space Exploration \\ Arizona State University, Tempe, AZ 85827-1404}

\begin{abstract}
In a recently proposed Higgs-Seesaw model the observed scale of dark energy results from a metastable false vacuum energy associated with mixing of the standard model Higgs particle and a scalar associated with new physics at the GUT or Planck scale.   Here we address the issue of how to ensure metastability of this state over cosmological time. We consider new tree-level operators, the presence of a thermal bath of hidden sector particles, and quantum corrections to the effective potential.  We find that in the thermal scenario many additional light degrees of freedom are typically required unless coupling constants are somewhat fine-tuned.  However quantum corrections arising from as few as one additional light scalar field can provide the requisite support.  We also briefly consider implications of late-time vacuum decay for the perdurance of observed structures in the universe in this model.  
\end{abstract}

\keywords{dark energy, Higgs, GUT}

\maketitle

\setlength{\parindent}{20pt}
\setlength{\parskip}{2.5ex}

\section{Introduction}\label{sec:Introduction}

Understanding the nature of dark energy, with an inferred magnitude of approximately $\rho_{\DE}^{\rm (obs)} = 28 \meV^4$ \cite{Sanchez:2012sg, Campbell:2012hi, Ade:2013lta}, remains the deepest open problem in particle physics and cosmology.  
Observations suggest that this source has an equation of state  $w = -1$, consistent with either a fundamental cosmological constant or 
false vacuum energy associated with a metastable scalar field configuration.  
In either case, quantum effects would suggest that this energy, $\rho_{\DE}$, will depend sensitively on unknown UV physics, and it is therefore very difficult to imagine how the observed small energy scale could naturally arise \cite{Weinberg:1988cp}.  In particular  (i) Why not $\rho_{\DE} = \Lambda^4$ where $\Lambda$ is the UV cutoff of the effective field theory, (ii) Why not a natural value $\rho_{\DE} = 0$, which could result from some symmetry constraint?  

The answers to these fundamental questions will most likely require an understanding of a full quantum theory of gravity.  Assuming they are resolvable, and that the ultimate vacuum energy is indeed zero, one can proceed to consider whether plausible physics, based on known energy scales in particle theory, might produce at least a temporary residual vacuum energy consistent with current observations.  
Recently in \rref{Krauss:2013oea} it was proposed   that a Higgs portal, mixing electroweak and grand unification $M_{GUT}$ scalars, might naturally produce the observed magnitude of the energy density of dark energy due to the false vacuum energy associated with an otherwise new massless scalar field that is a singlet under the SM gauge group.     The questions we examine here are whether it is possible to ensure that this field remains in its false vacuum state for cosmological times, and what the implications might be for the future when it decays to its true ground state.

The organization of this paper is as follows.  
In \sref{sec:Review} we review the Higgs-Seesaw model of dark energy, and in particular, we estimate the lifetime of the false vacuum in this model.  In \sref{sec:Support} we explore three variants of the minimal model that extend the lifetime of the false vacuum to cosmological time scales.  
Since the false vacuum is only metastable, it will eventually decay, and we consider the implications of this decay in \sref{sec:Implications}.  
We conclude in \sref{sec:Conclusions}.

\section{Review of the Higgs-Seesaw Model}\label{sec:Review}

The model of \rref{Krauss:2013oea} extends the SM by introducing a complex scalar field $\sigma_{\HS}$, which is a singlet under the SM gauge groups and charged under its own global axial symmetry.  
Denoting the SM Higgs doublet as $\Phi_{\SM} = ( \phi_{\SM}^{+} \, , \, \phi_{\SM})^T$, the scalar sector Lagrangian is written as 
\begin{align}
	\Lcal = \bigl| \partial_{\mu} \Phi_{\SM} \bigr|^2 + \abs{\partial_{\mu} \sigma_{\HS}}^2 - V(\Phi_{\SM} , \sigma_{\HS})
\end{align}
where
\begin{align}\label{eq:V}
	V(\Phi_{\SM}, \sigma_{\HS}) = \Omega_{\CC} + \mu^2 \abs{\Phi_{\SM}}^2 +  \lambda \abs{\Phi_{\SM}}^4 + \lambda_{\rm mix} \abs{\Phi_{\SM}}^2 \abs{\sigma_{\HS}}^2 + \lambda_{\HS} \abs{\sigma_{\HS}}^4 \per
\end{align}
The bi-quadratic term is sometimes referred to as the Higgs portal operator \cite{Silveira:1985rk, Patt:2006fw}.  
If this operator arises by virtue of GUT-scale physics, as argued in \rref{Krauss:2013oea}, then its value should naturally be extremely small in magnitude
\begin{align}\label{eq:lamnat}
	\lambda_{\rm mix}^{\rm (nat)} \approx \frac{M_W^2}{M_{GUT}^2} \simeq 6.5 \times 10^{-29} \left( \frac{M_{GUT}}{10^{16} \GeV} \right)^{-2} \per
\end{align}
Note the absence of a mass term for the field $\sigma_{\HS}$, which is assumed, due to symmetries in the GUT-scale sector, to only acquire a mass after electroweak symmetry breaking.   

For the purposes of studying the vacuum structure it is convenient to take $\phi_{\SM}^{+} = 0$, $\phi_{\SM} = h / \sqrt{2}$, and $\sigma_{\HS} = s / \sqrt{2}$ where $h(x)$ and $s(x)$ are real scalar fields.  
Then the scalar potential becomes
\begin{align}\label{eq:Uhs}
	U(h,s) = \Omega_{\CC} + \frac{1}{2} \mu^2 h^2 + \frac{\lambda}{4} h^4 + \frac{\lambda_{\rm mix}}{4} h^2 s^2 + \frac{\lambda_{\HS}}{4} s^4 \per
\end{align}
where $\Omega_{\CC}$ is a bare cosmological constant which must be tuned to cancel UV contributions from the scalar field sector. The tachyonic mass $\mu^2 = - \lambda v^2$ induces electroweak symmetry breaking and causes the Higgs field to acquire a vacuum expectation value $\langle h \rangle = v$, which in turn induces a mass $\mu_{\HS}^2 = \lambda_{\rm mix} v^2 / 2$ for the field $s$.  
If $\lambda_{\rm mix} < 0$ then this mass is tachyonic, and the true vacuum state of the theory is displaced to 
\begin{align}
	\langle h \rangle_{\rm true} & \equiv v_{h} = \frac{v}{\sqrt{1- \epsilon^2}} \approx v \left[ 1 + O(\epsilon^2) \right] \label{eq:htrue} \\
	\langle s \rangle_{\rm true} & \equiv v_{s} = v \left( \frac{\lambda}{\lambda_{\HS}} \right)^{1/4} \sqrt{ \frac{\epsilon}{1 - \epsilon^2} } \approx v \left( \frac{\lambda}{\lambda_{\HS}} \right)^{1/4} \sqrt{\epsilon} \  \Bigl[ 1 + O(\epsilon^{2}) \Bigr] \label{eq:strue} 
\end{align}
where $\epsilon \equiv - \lambda_{\rm mix} / \sqrt{4 \lambda \lambda_{\HS}}$.  
We will use $\langle h \rangle_{\rm false} = v$ and $\langle s \rangle_{\rm false} = 0$ to denote the tachyonic false vacuum state.  

For typical values of the coupling $\lambda_{\rm mix}$, see \eref{eq:lamnat}, the mass scales of the $\sigma_{\HS}$ field are extremely small:  $\mu_{\HS}^2 \lll \mu^2$ and $v_s \lll v_h \sim v$.  
In this limit it is a good approximation to integrate out the Higgs field and work with an effective field theory for the $\sigma_{\HS}$ field alone.  
The field equation $\partial U / \partial h = 0$ has the solution
\begin{align}
	\bar{h}(s) = v \sqrt{1 + \epsilon \left(\frac{\lambda_{\HS}}{\lambda}\right)^{1/2} \, s^2 } \com
\end{align}
which interpolates between the false and true vacua, as one can easily verify.  
The scalar potential in the effective theory, $U(s) \equiv U(\bar{h}(s),s)$, is given by 
\begin{align}\label{eq:Us}
	U(s) = \left( \Omega_{\CC} - \frac{\lambda v^4}{4} \right) + \frac{\lambda_{\rm mix}}{4} v^2 s^2 + \frac{\lambda_{\HS}}{4} \left( 1 - \epsilon^2 \right) s^4 \per
\end{align}

If it is assumed that the scalar potential vanishes in the true vacuum, {\it i.e.} $U(v_s) = 0$, then the bare cosmological constant must be tuned to be 
\begin{align}\label{eq:tuning}
	\Omega_{\CC} 
	= \frac{\lambda v^4}{4} \frac{1}{1 - \epsilon^2} 
	\approx \frac{\lambda v^4}{4} \left[ 1 + \epsilon^2 + O(\epsilon^4) \right] \per
\end{align}
The effective cosmological constant today will then be smaller than $\Omega_{\CC}$ as a consequence of symmetry breaking phase transitions.  
If the scalar fields have not reached their true vacuum state but are instead suspended in the false vacuum, then the vacuum energy density, $\rho_{\DE} \equiv U(0)$, is given by 
\begin{align}\label{eq:rhoDE_analytic}
	\rho_{\DE} 
	= \frac{\lambda v^4}{4} \frac{\epsilon^2}{1-\epsilon^2}
	\approx \frac{\lambda_{\rm mix}^2 v^4}{16 \lambda_{\HS}} \per
\end{align}
As the notation suggests, $\rho_{\DE}$ should be identified with the energy density of dark energy.  
Taking $\lambda = M_H^2 / (2 v^2)$ with $v \simeq 246 \GeV$ and $M_H \simeq 126 \GeV$ gives 
\begin{align}\label{eq:rhoDE_numerical}
	\rho_{\DE} \simeq 0.97 \meV^4 \left( \frac{\lambda_{\rm mix}}{\lambda_{\rm mix}^{\rm (nat)}} \right)^2 \left( \frac{1}{\lambda_{\HS}} \right) \per
\end{align}
This value is comparable to the observed energy density, $\rho_{\DE}^{\rm (obs)} \approx 28 \meV^4$ \cite{Ade:2013lta}.  
In this way, the Higgs-Seesaw model naturally predicts the correct magnitude for the energy density of dark energy density from the electroweak and GUT scales.  
For the discussion in the following sections, it will be useful here to rewrite \eref{eq:rhoDE_numerical} as 
\begin{align}\label{eq:lamHS_numerical}
	\lambda_{\HS} \simeq 0.035 \left( \frac{\lambda_{\rm mix}}{\lambda_{\rm mix}^{\rm (nat)}} \right)^2 \left( \frac{\rho_{\DE}^{\rm (obs)}}{\rho_{\DE}} \right) 
\end{align}
and to note that $\lambda_{\HS}$ remains perturbatively small for $\lambda_{\rm mix} \leq \lambda_{\rm mix}^{\rm (nat)}$.  

The success of the Higgs-Seesaw model hinges upon the assumption that the universe is trapped in the false vacuum.  
The lifetime of the false vacuum can be estimated by dimensional analysis using the tachyonic mass scale, $\mu_{\HS}^2 = \lambda_{\rm mix} v^2/2$.  
Taking the same numerical values as above, this time scale is 
\begin{align}\label{eq:tachyon_numerical}
	\abs{\mu_{\HS}}^{-1} \simeq (1.4 \meV)^{-1} \approx 0.47 \, {\rm nanoseconds} \, .
\end{align}
Therefore, in the absence of any support, the false vacuum would have decayed in the very early universe.  
This observation motivates the present work, in which we will explore scenarios that can provide support to the tachyonic false vacuum, following a classification scheme outlined in \rref{Chung:2012vg}

\section{ Three Support Mechanisms}\label{sec:Support}

\subsection{Tree-Level Support}\label{sub:Tree}

The presence of additional terms in the tree-level scalar potential, $V(\Phi_{\SM}, \sigma_{\HS})$, can provide support for the tachyonic false vacuum.   As we now demonstrate, this option does not appear viable however. 

The most straightforward way to lifting the tachyonic instability is to add a mass term,
\begin{align}
	\delta V = m_{\HS}^2 \abs{\sigma_{\HS}}^2 \com
\end{align}
such that $m_{\HS}^2 + \mu_{\HS}^2 > 0$.  
Forgetting for the moment, the question of what is the natural scale for $m_{\HS}$, we can consider the implications of adding such a term to the potential.  
Not only does this term succeed in lifting the tachyon, it additionally changes the vacuum structure of the theory in such a way that the false vacuum becomes absolutely stable.  
Since the Higgs-Seesaw dark energy model assumes that the true vacuum state has a vanishing vacuum energy density, this implies that the cosmological constant should vanish in our universe today, {\it i.e.} $\rho_{\DE} = 0$, which is unacceptable.  

Alternatively, we can extend the potential by the non-renormalizable operator
\begin{align}\label{eq:V_nonren}
	\delta V = \frac{\abs{\sigma_{\HS}}^6}{M^2} \per
\end{align}
In the context of Higgs-Seesaw dark energy, the SM is understood to be an effective field theory with a cutoff at the GUT scale, $M_{GUT} \sim 10^{16} \GeV$.  
Therefore, the natural choice for the parameter $M$ is $M \sim M_{GUT}$.  
Upon added $\delta V$, the scalar potential becomes
\begin{align}\label{eq:U_nonren}
	U(s) = \left( \Omega_{\CC} - \frac{\lambda}{4} v^4 \right) + \frac{\lambda_{\rm mix}}{4} v^2 s^2 + \frac{\lambda_{\HS}}{4} \left[ 1 - \frac{\lambda_{\rm mix}^2}{4 \lambda \lambda_{\HS}} \right] s^4 + \frac{s^6}{8 M^2} \per
\end{align}
By choosing $\lambda_{\rm mix} > 0$ and $\lambda_{\HS} < 0$, we have a potential\footnote{A scalar potential of this form has been studied in the context of the electroweak phase transition \cite{Grojean:2004xa, Delaunay:2007wb, Grinstein:2008qi}.  } with a metastable minimum at $\langle s \rangle_{\rm false} = 0$ and an absolute minimum at 
\begin{align}
	\langle s \rangle_{\rm true} 
	= \sqrt{ \frac{2}{3} } M \sqrt{-\lambda_{\HS}} \sqrt{1 + \bar{\epsilon}^2} \sqrt{1 + \sqrt{1 + \tilde{\epsilon} }}
	\approx \frac{2}{\sqrt{3}} M \sqrt{- \lambda_{\HS}} \bigl[ 1 + O(\bar{\epsilon}, \tilde{\epsilon}) \bigr]
\end{align}
where $\bar{\epsilon} \equiv \lambda_{\rm mix} / \sqrt{-4 \lambda \lambda_{\HS}}$ and $\tilde{\epsilon} \equiv 3 \lambda_{\rm mix} v^2/ [2 \lambda_{\HS}^2 M^2(1+\bar{\epsilon}^2)^2]$.  
The VEV of $s$ in the true vacuum is now set by the cutoff scale $M$, and not by the small quantity $\sqrt{\epsilon} \, v$ as in \eref{eq:strue}.  
Similarly, we find that the false vacuum is lifted above the true vacuum by an energy density
\begin{align}
	\Delta \rho = U(0) - U(\langle s \rangle_{\rm true}) = \frac{4}{27} (-\lambda_{\HS})^3 M^4 \bigl[ 1 + O(\epsilon) \bigr] \per
\end{align}
For the natural GUT scale cutoff, this quantity is many orders of magnitude larger than the observed energy density of dark energy unless $\lambda_{\HS} \lll 1$.   


\subsection{Thermal Support}\label{sub:Thermal}

Symmetry restoration can also result as a consequence of thermal effects.  
In the Standard Model, the tachyonic mass of the fundamental Higgs field, $\mu^2 = - M_H^2 /2 \simeq -(89 \GeV)^2$, is lifted by thermal corrections when the temperature of the universe exceeds $T = T_c \approx 150 \GeV$ \cite{Anderson:1991zb}.  
The relationship $T_c = O(\abs{\mu_{\HS}})$ is a result of dimensional analysis.  
By analogy, one will naively expect the tachyonic mass scale of the Higg-Seesaw model to be lifted at temperatures $T \gtrsim \abs{\mu_{\HS}} \approx 1.4 \meV$ [see \eref{eq:tachyon_numerical}].  
For reference, the current temperature of the CMB is $T_{\rm cmb} \approx 2.73 \, {\rm Kelvin} \approx 0.23 \meV$.  
The observation that $\abs{\mu_{\HS}} > T_{\rm cmb}$ is our first indication that the thermal support scenario will be a difficult to implement in a phenomenologically viable way.  
In order for the thermal support scenario to be successful, we will see that the minimal Higgs-Seesaw model must be extended to include a thermal bath of many new light fields coupled to the tachyonic field $\sigma_{\HS}$.  

Assume then that the universe is permeated by a thermal bath of such relativistic, hidden sector (HS) particles\footnote{If they are not relativistic, their contribution to the thermal mass correction is Boltzmann suppressed and therefore negligible.}.  
For concreteness we will assume that these particles are scalars, but they could just as well have higher spin and our analysis would be qualitatively unchanged.  
For the sake of generality, suppose that there are $N_s$ distinct species of scalar particles with a common temperature $T_{\HS}$, with $g_{\ast \HS} = N_s$ effective degrees of freedom, and with an energy density
\begin{align}\label{eq:rhoHS}
	\rho_{\HS} = \frac{\pi^2}{30} g_{\ast \HS} T_{\HS}^4 \per
\end{align}
The relativistic energy density of the universe is constrained via the CMB; the constraints are quoted in terms of the ``effective number of neutrinos'' $N_{\rm eff} \approx 3.30 \pm 0.27$ \cite{Ade:2013lta}.  
The energy density of the HS thermal bath yields a contribution 
\begin{align}\label{eq:Neff}
	\Delta N_{\rm eff} \equiv \frac{\rho_{\HS}}{2 \frac{\pi^2}{30} \frac{7}{8} T_{\nu}^4} 
	= \frac{4N_s}{7} \left( \frac{T_{\HS}}{T_{\nu}} \right)^4 \per
\end{align}
where $T_{\nu} \approx 1.7 \times 10^{-4} \eV$ is the temperature of the cosmic neutrino background today.  
The task herein is to determine if this gas can be hot enough and have enough degrees of freedom in order to stabilize the tachyonic field while also keeping its energy density low enough to satisfy the empirical constraint $\Delta N_{\rm eff} \lesssim O(1)$.  

To be concrete, let us denote the new, real scalar fields as $\varphi_i(x)$ and suppose that they couple to the tachyonic field $\sigma_{\HS}$ through the interaction 
\begin{align}\label{eq:dV}
	\delta V \ni \abs{\sigma_{\HS}}^2 \sum_{i=1}^{N_s} g_{\varphi_i}^2 (\varphi_i)^2 
\end{align}
where $g_{\varphi_i}^2$ are the coupling constants.  
This interaction gives rise to the thermal mass correction \cite{Kapusta:1989}
\begin{align}\label{eq:muth}
	\mu_{th}^2 \approx \frac{N_s g_{\varphi}^2 T_{\HS}^2}{12} 
\end{align}
where we have defined $g_{\varphi}^2 \equiv (N_s)^{-1} \sum_{i=1}^{N_s} g_{\varphi_i}^2$.  
It is important to recognize that the powers of $N_{s}$ and $T_{\HS}$ in Eqs.~(\ref{eq:Neff})~and~(\ref{eq:muth}) are not the same.  
By decreasing $T_{\HS}$ and scaling $N_{s} \sim 1 / T_{\HS}^4$, we can keep $\Delta N_{\rm eff} < O(1)$ while increasing $\mu_{th}^2 \sim 1 / T_{\HS}^2$.  

\begin{figure}[t]
\begin{center}
\includegraphics[width=0.70\textwidth]{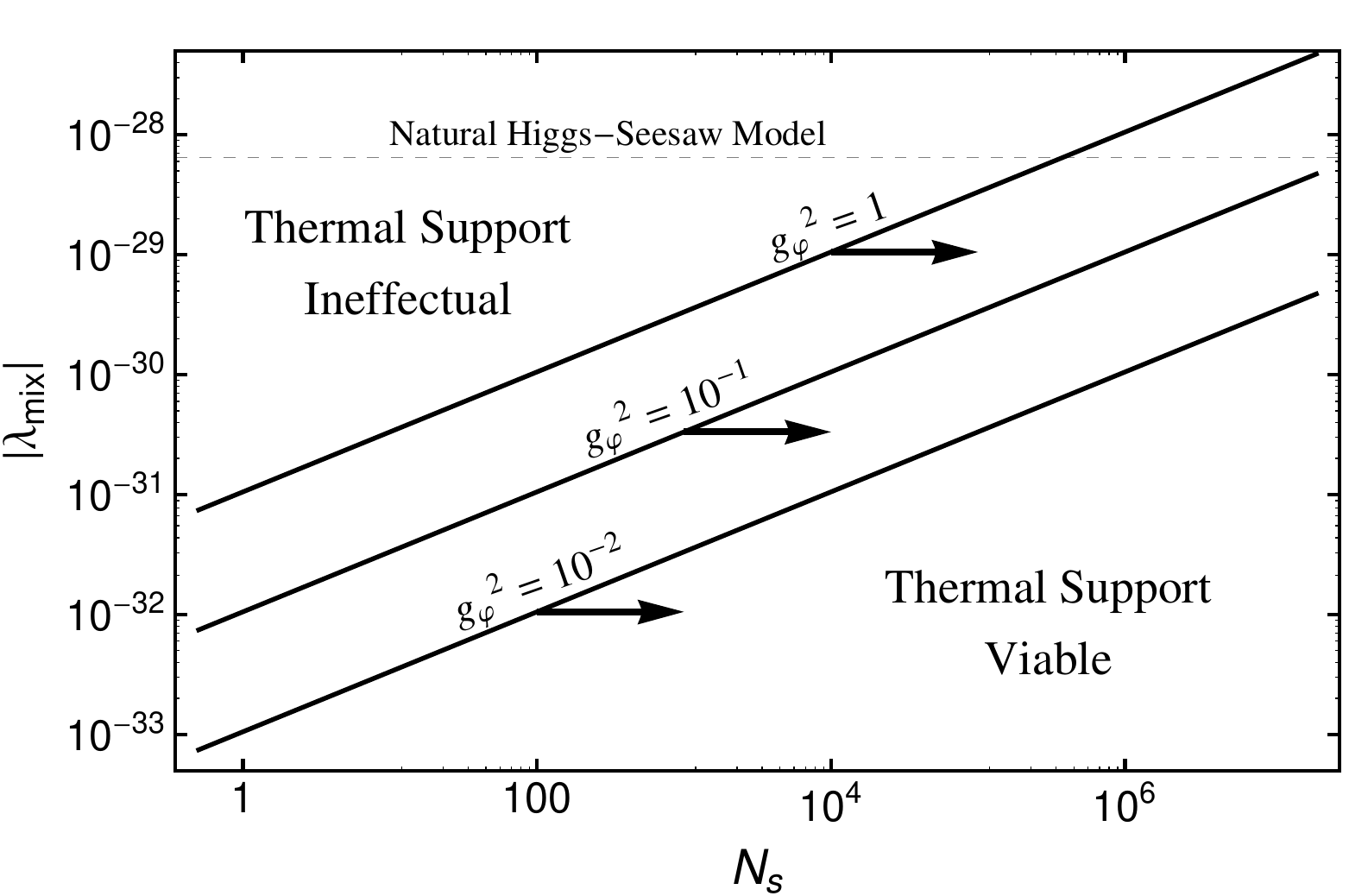} 
\caption{
\label{fig:thermsupport}
The constraint \eref{eq:lammix_range}.  For a given value of $g_{\varphi}^2$ (decreasing from top to bottom) the solid black lines demarcate a boundary to the right of which thermal support is a viable means of lifting the tachyonic instability provided that the temperature of the hidden sector thermal bath, $T_{\HS}$, falls in the range given by \eref{eq:THS_range}.  
}
\end{center}
\end{figure}

\begin{figure}[t]
\begin{center}
\includegraphics[width=0.95\textwidth]{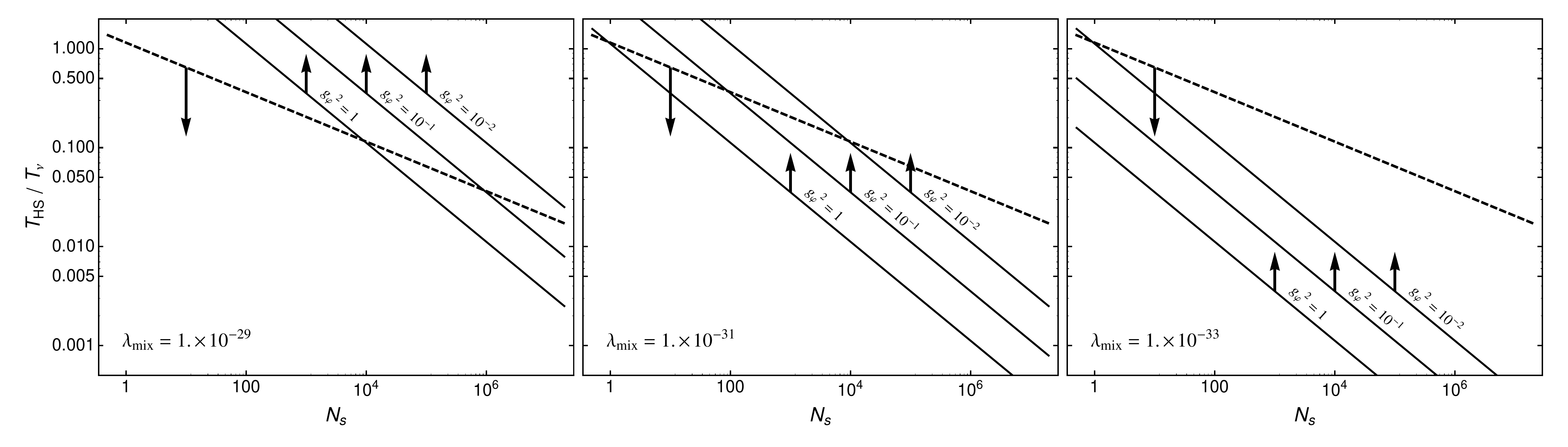} 
\caption{
\label{fig:Tplot}
For a given $N_s$, thermal support is viable over the range of temperatures bounded above by the dashed line and bounded below by one of the solid lines (for $g_{\varphi}^2 = 10^{-2}, 10^{-1}, 1$ from top to bottom).  See \eref{eq:THS_range}.  
}
\end{center}
\end{figure}

As we discussed above, we want to determine the range of parameters for which the bounds 
\begin{align}
	\mu_{th}^2 > \abs{\mu_{\HS}^2} 
	\qquad {\rm and} \qquad
	\Delta N_{\rm eff} < 1
\end{align}
are satisfied.  
These constraints can be resolved as 
\begin{align}\label{eq:lammix_range}
	\abs{ \lambda_{\rm mix} }
	&< \frac{\sqrt{7}}{12} \frac{g_{\varphi}^2 \sqrt{N_s} T_{\nu}^2}{v^2} 
	\approx 1.6 \  \lambda_{\rm mix}^{\rm (nat)} \left( \frac{g_{\varphi}^2}{1} \right) \left( \frac{N_s}{10^{6}} \right)^{1/2}
\end{align}
and
\begin{align}\label{eq:THS_range}
	\sqrt{\frac{6 \abs{\lambda_{\rm mix}}}{g_{\varphi}^2 N_s} } v 
	<
	T_{\HS} 
	&< \left( \frac{7}{4 N_s} \right)^{1/4} T_{\nu} \per
\end{align}
Note that when the bound in \eref{eq:lammix_range} is saturated, the range in \eref{eq:THS_range} vanishes.  
The bound in \eref{eq:lammix_range} is shown in \fref{fig:thermsupport}.  
For the natural Higgs-Seesaw model, $\lambda_{\rm mix} = O(\lambda_{\rm mix}^{\rm (nat)})$, the number of new scalar degrees of freedom must be very large, $N_s \sim 10^{6}$.  
Conversely, if thermal support is to be established using only $N_s = O(1)$ new scalar degrees of freedom then the coupling must be smaller, $\lambda_{\rm mix} \approx 10^{-3} \lambda_{\rm mix}^{\rm (nat)}$ for $g_{\varphi}^2 = 1$.  
Decreasing $g_{\varphi}^2$ makes these constraints more stringent.  
For a given point in the parameter space represented in \fref{fig:thermsupport}, the temperature of the thermal bath, $T_{\HS}$, must satisfy \eref{eq:THS_range}; these constraints are shown in \fref{fig:Tplot}.  
Although $T_{\HS}$ can be as large as $1.15 T_{\nu}$, its value is typically smaller by one or two orders of magnitude over the parameter space shown here.  

\subsection{Loop-Level Support}\label{sub:Loop}

As a final example, we consider the role that quantum corrections to the effective potential can play.  
In particular, we will consider a potential with the structure $U \sim + s^2 - s^4 + s^4 \ln s^2$ where the quadratic term has a positive coefficient provided by taking $\lambda_{\rm mix} > 0$, the quartic term has a negative coefficient, and the logarithmic term arises from one-loop quantum corrections of the Coleman-Weinberg form \cite{Coleman:1973jx}.  
To obtain the appropriate quantum corrections, the model must be extended to include additional {\it scalar} fields \cite{Espinosa:2007qk}, since fermionic field would yield quantum corrections with the wrong sign.  

We consider the same extension of the Higgs-Seesaw model that was discussed in \sref{sub:Thermal}.  
Namely, we introduce $N_s$ real scalar fields $\varphi_i$ that coupled to $\sigma_{\HS}$ through the interaction given previously by \eref{eq:dV}.  
In the presence of a background field $\langle \sigma_{\HS} \rangle = s / \sqrt{2}$, the new scalars acquire masses
\begin{align}
	m_{\varphi_i}(s) = g_{\varphi_i} s \per
\end{align}
For simplicity we will assume a universal coupling $g_{\varphi_i} = g_{\varphi}$.  
The renormalized one-loop effective potential is given by $V_{\rm eff}(s) = U(s) + V_{\CW}(s)$
where $U$ was given previously by \eref{eq:Us}, and the Coleman-Weinberg potential is \cite{Coleman:1973jx}
\begin{align}
	V_{\CW}(s) = \sum_{i=1}^{N_s} \frac{m^{4}_{\varphi_i}(s)}{64 \pi^2} \left( \ln \frac{m^2_{\varphi_i}(s)}{M^2} - \frac{3}{2} \right) \com
\end{align}
after employing dimensional regularization and renormalizing in the $\overline{\rm MS}$ scheme at a scale $M$.  

Since we seek to study the issue of vacuum stability, it is useful at this point to exchange some of the parameters in $V_{\rm eff}$ in favor of parameters with a more direct relevance to the vacuum structure.  
We will focus on models for which $V_{\rm eff}(s)$ has a global minimum at $s = v_s$ with $V_{\rm eff}(v_s) = \rho_{\rm true} = 0$ and a local minimum at $s = 0$ with $V_{\rm eff}(0) = \rho_{\DE} + \rho_{\rm true}$.  
Thus, we exchange the parameters $\Omega_{\CC}$ and $\lambda_{\HS}$ in favor of $\rho_{\rm true}$ and $\rho_{\DE}$.  
Taking the renormalization scale to be $M = g_{\varphi} v_s$, the effective potential becomes
\begin{align}\label{eq:Veff_loop}
	V_{\rm eff}(s) = \rho_{\DE} 
	+ \frac{\lambda_{\rm mix}}{4} v^2 s^2 
	- \frac{1}{4} \left( \lambda_{\rm mix} \frac{v^2}{v_s^2} + 4 \frac{\rho_{\DE} - \rho_{\rm true}}{v_s^4} \right) s^4
	+ \frac{N_s g_{\varphi}^4 s^4}{64 \pi^2} \ln \frac{s^2}{v_s^2}  \per
\end{align}
where
\begin{align}\label{eq:vs_def}
	v_s = 2 \sqrt{2} \pi v \sqrt{ \frac{\lambda_{\rm mix}}{N_s g_{\varphi}^4} } \sqrt{1 + \sqrt{ 1 + \frac{2N_s g_{\varphi}^4 (\rho_{\DE} - \rho_{\rm true})}{\pi^2 \lambda_{\rm mix}^2 v^4} } } \per
\end{align}
As promised, $V_{\rm eff}$ has the structure ``$+s^2-s^4+s^4\ln s^2$'' provided that $\lambda_{\rm mix} > 0$.  

\begin{figure}[t]
\begin{center}
\includegraphics[width=0.99\textwidth]{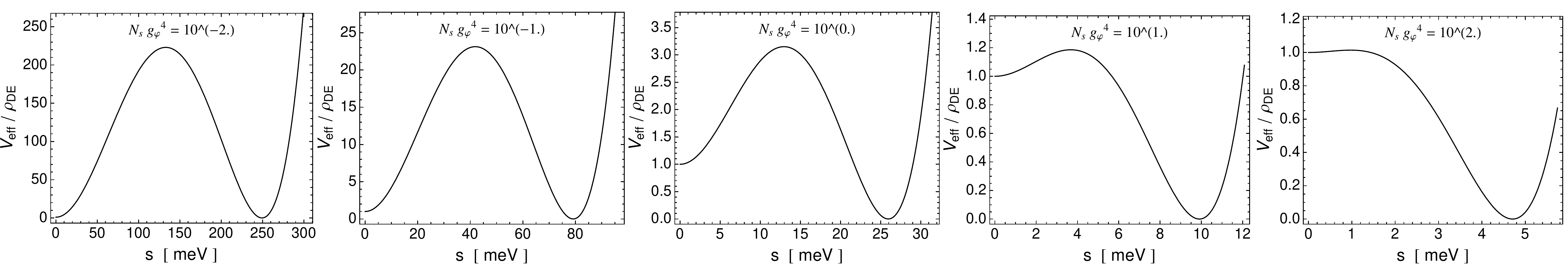} 
\caption{
\label{fig:Veff_loop}
The effective potential, \eref{eq:Veff_loop}, for $\lambda_{\rm mix} = \lambda_{\rm mix}^{\rm (nat)} \approx 6.5 \times 10^{-29}$, $\rho_{\rm true} = 0$, $\rho_{\DE} = \rho_{\DE}^{\rm (obs)} \approx 28 \meV^4$, and a range of values for $N_s g_{\varphi}^4$.  Note the axes are scaled differently in the the different panels.  
}
\end{center}
\end{figure}

This effective potential is shown in \fref{fig:Veff_loop} for $\lambda_{\rm mix} = \lambda_{\rm mix}^{\rm (nat)}$ and various values of $N_s g_{\varphi}^4$.  
The false vacuum ($s = 0$) is always metastable thanks to the quadratic term in \eref{eq:Veff_loop}.  
Asymptotically the barrier height increases in the limit $N_s g_{\varphi}^4 \to 0$, and it decreases as $N_s g_{\varphi}^4 \to \infty$.  
This somewhat counterinuitive behavior is a consequence of the way that we allow $\lambda_{\HS} \sim 3 N_s g_{\varphi}^4 / 32 \pi^2$ to vary such that $\lambda_{\rm mix}$ and $\rho_{\DE} - \rho_{\rm true}$ can remain fixed.  
The crossover between the large and small barrier regimes occurs when $N_s g_{\varphi}^4 = O(1)$.  
To understand this better, we can approximate the barrier height as $V_{\rm barrier} \approx V_{\rm eff}(s=v_s/2)$, which gives (also setting $\rho_{\rm true} = 0$)
\begin{align}\label{eq:Vbarrier}
	\frac{V_{\rm barrier}}{\rho_{\DE}} \approx \frac{15}{16} + \frac{3}{64} \frac{\lambda_{\rm mix} v^2 v_s^2}{\rho_{\DE}} - \frac{1}{1024 \pi^2} \frac{N_s g_{\varphi}^4 v_s^4 \ln 4}{\rho_{\DE}} \per
\end{align}
Recall that $v_s$ was given by \eref{eq:vs_def}.  If we define the factor
\begin{align}\label{eq:kappa}
	\kappa \equiv \frac{2 N_s g_{\varphi}^4 \rho_{\DE}}{\pi^2 \lambda_{\rm mix}^2 v^4}
\end{align}
then in limits of small and large $N_s g_{\varphi}^4$, the barrier height is approximated as 
\begin{align}\label{eq:Vbarrier_limits}
	\frac{V_{\rm barrier}}{\rho_{\DE}} \approx 
	\begin{cases}
	\frac{1}{N_s g_{\varphi}^4 } \frac{\pi^2 (3 - \ln 4)}{4} \frac{ \lambda_{\rm mix}^2 v^4}{\rho_{\DE}} & \kappa \ll 1\\
	\frac{15 - \ln 16}{16} & \kappa \gg 1
	\end{cases} \per
\end{align}
The crossover occurs when $\kappa = O(1)$, and since $\rho_{\DE}^{\rm (obs)} \sim (\lambda_{\rm mix}^{\rm (nat)})^2 v^4$, as we saw in \eref{eq:rhoDE_analytic}, this corresponds to $N_s g_{\varphi}^4 = O(\lambda_{\rm mix}^2 / \lambda_{\rm mix}^{\rm (nat) \, 2})$.  
In other words, for $\lambda_{\rm mix} \ll \lambda_{\rm mix}^{\rm (nat)}$ the barrier height is very small when $N_s g_{\varphi}^4 = O(1)$, and for $\lambda_{\rm mix} \gg \lambda_{\rm mix}^{\rm (nat)}$ the barrier is large for $N_s g_{\varphi}^4 = O(1)$.  

\begin{figure}[t]
\begin{center}
\includegraphics[width=0.70\textwidth]{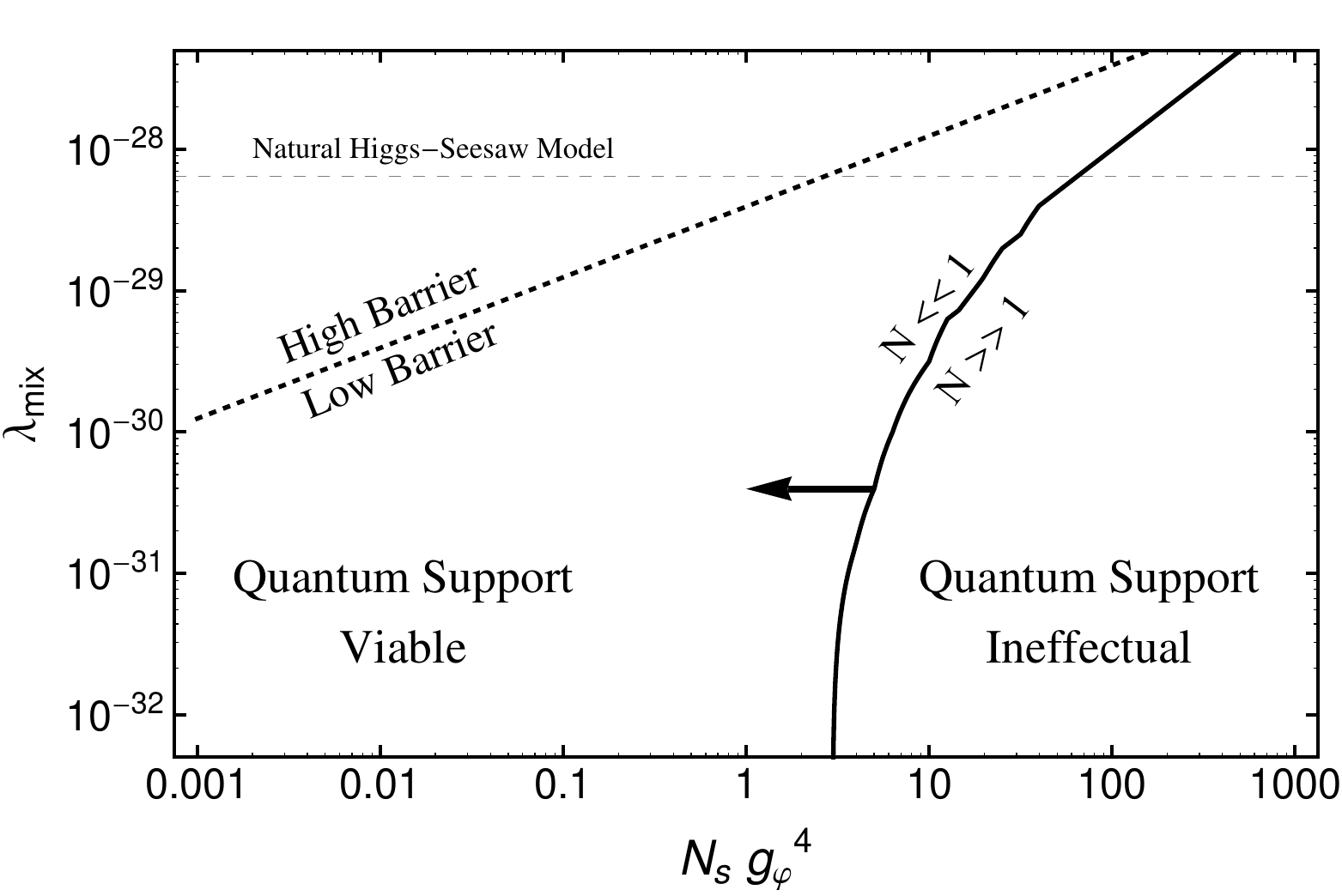} 
\caption{
\label{fig:param_loop}
The parameter space of the quantum support scenario.  Along the dotted line the parameters satisfy $\kappa = 1$ [see \eref{eq:kappa}], and along the solid line they satisfy $\mathcal{N} = 1$ [see \eref{eq:Nbub}].  
}
\end{center}
\end{figure}

The metastable vacuum ($s =0$) can decay via quantum tunneling.  
Using standard techniques \cite{Coleman:1977py, Callan:1977pt} we calculate the Euclidean action $B$ of the tunneling solution and evaluate the decay rate per unit volume as 
\begin{align}\label{eq:GammaV}
	(\Gamma/V) = \mu_{\HS}^4 \frac{B^2}{4 \pi^2} e^{-B}
\end{align}
where $\mu_{\HS}^2 = \lambda_{\rm mix} v^2 / 2$.  
The number of bubble nucleation events integrated over a Hubble volume ($d_H^3 \approx H_0^{-3}$) and the age of the universe ($t_U \approx H_0^{-1}$) is then estimated as 
\begin{align}\label{eq:Nbub}
	\mathcal{N} = (\Gamma/V)H_0^{-4}
\end{align}
where $H_0 \simeq 1.5 \times 10^{-33} \eV$ is the Hubble constant today.  
The quantum support scenario is viable when $\mathcal{N} \ll 1$ and ineffectual when $\mathcal{N} \gg 1$.  
In \fref{fig:param_loop} we show the viability of quantum support over the parameter space.  
The dotted line demarcates the threshold between the small and large barrier regimes ($\kappa = 1$), and the solid line indicates the boundary between viable and ineffectual quantum support ($\mathcal{N} = 1$).  
As $\lambda_{\rm mix} \to 0$, the condition $\mathcal{N} =1$ becomes independent of $\lambda_{\rm mix}$ as a result of the way in which we scale $\lambda_{\HS}$ in order to hold $\rho_{\DE}$ fixed.\footnote{
The bounce action is $B = 2 \pi^2 \int r^3 dr \left[ 1/2 (d s / d r)^2 + V_{\rm eff}(s) \right]$.  After rescaling $s = v_s \bar{s}$, $V_{\rm eff} = \rho_{\DE} \bar{V}_{\rm eff}$, and $r = (v_s / \sqrt{\rho_{\DE}}) \bar{r}$ it becomes $B = (v_s^4 / \rho_{\DE}) \bar{B}$ where $\bar{B}$ predominantly depends on the ``shape'' of the effective potential, particularly the height of the barrier relative to the scale of degeneracy breaking.  
As we saw in \eref{eq:Vbarrier_limits}, $V_{\rm barrier} / \rho_{\DE}$ becomes independent of all parameters in the limit $\kappa \gg 1$, which corresponds to $\lambda_{\rm mix} \to 0$.  
Additionally, the prefactor $v_s^4 / \rho_{\DE} \approx 128 \pi^2 / (N_s g_{\varphi}^4) + O(\lambda_{\rm mix}^2)$ is independent of $\lambda_{\rm mix}$ to leading order.  
Thus, the asymptotic behavior seen in \fref{fig:param_loop} is explained.  
}
For the natural parameter choice, $\lambda_{\rm mix} \approx \lambda_{\rm mix}^{\rm (nat)}$ and $N_s g_{\varphi}^4 = O(1)$, the metastable vacuum has a lifetime that exceeds the age of the universe.

\section{Implications of Late-Time Vacuum Decay}\label{sec:Implications}

While various simple extensions of the Higgs-Seesaw model appear to make it possible for the false vacuum to be supported over cosmological time intervals, nevertheless, the eventual decay of the false vacuum is inescapable.   In the case of thermal support, the thermal bath will eventually cool due to the expansion of the universe, and the tachyonic instability will reemerge.  
For typical parameters (see \fref{fig:Tplot}), the allowed temperature range of the hidden sector gas spans only one or two decades.  Thus in this case we would expect thermal support to be lost within the next few Hubble times once the hidden sector gas cools sufficiently.   This makes the possibility that our vacuum could decay in the not-too-distant future somewhat less fine-tuned than in the quantum support case, where the lifetime of the false vacuum is exponentially sensitive to the parameters, and the metastable state may be extremely long lived if $\mathcal{N} \lll 1$ (see \fref{fig:param_loop}).   Either way, the false vacuum will eventually decay.

It is worth mentioning, at least briefly that it is possible in principle that the false vacuum has already decayed and that we now sit in the true vacuum with vanishing vacuum energy \cite{Goldberg:2000ap, delaMacorra:2007pw, Dutta:2009ix, Abdalla:2012ug, Pen:2012hn}.  However, such a possibility is extremely remote as it requires extreme fine tuning.  In this case, in general the universe today would now be radiation dominated, which is ruled out unless the decay occurred extremely recently (i.e. see \cite{Turner:1984nf} ) and thus we shall not consider it further here.

A much more interesting question is what `observable' effects would result from future decay of the false vacuum in this model.  The word observable is unusual here because in general one might expect that a change in vacuum state would be a catastrophic process for the spectrum of particles and fields, and hence for all structures that currently exist.  

However, there is good reason to believe that this would not be the case in this model.  The primary effect of a change in the vacuum in this case would be a small shift in the VEV and couplings of the standard model Higgs field, to which the singlet scalar would become mixed.   However this effect is of the order of 
\begin{equation}
	\frac{\langle h \rangle_{\rm true} - v}{v} = O\left( \lambda_{\rm mix}^2 \right)
\end{equation}
[see \eref{eq:htrue}] and therefore will result in changes in elementary particle masses by less than $O(10^{-57})$.  It is hard to imagine that such a shift would produce any instability in bound systems of quarks, nucleons, or atoms. ( It also implies, for the same reason, that no terrestrial experiment we could perform on the Higgs at accelerators could in fact determine if this decay has already occurred. )

At the same time, the energy density stored in the false vacuum, while dominant in a cosmological sense, is subdominant on all scales smaller than that of clusters.  And while the release of this energy into relativistic particles might otherwise unbind the largest clustered systems, all such systems would already be unbound due to the expansion induced by the currently observed dark energy.

We therefore may be living in the best of all possible worlds, namely one in which the observed acceleration of the universe that will otherwise remove all observed galaxies from our horizon \cite{Krauss:1999hj, Krauss:2007nt} will one day end, but also one in which galaxies, stars, planets, and lifeforms may ultimately still survive through a phase transition and persist into the far future.


\section{Conclusions}\label{sec:Conclusions}

Contrary to naive expectations perhaps, we have demonstrated that it is possible to stabilize a false vacuum associated with a Higgs-Seesaw model of dark energy, which naively has a lifetime of $O(10^{-9})$ seconds, so that false vacuum decay can be suppressed for periods in excess of $10^{10}$ years, without drastically altering the characteristics of the model, or destroying the natural scales inherent within it.   Moreover,in this case, even if we are living in a false vacuum, we need not fear for the future.

\begin{acknowledgments}
This work was supported by ANU and by the DOE under Grant No.\ DE-SC0008016.  
\end{acknowledgments}


\providecommand{\href}[2]{#2}\begingroup\raggedright\endgroup

\end{document}